\documentstyle[aps,prl,multicol,epsfig]{revtex}
\begin{document}
\draft 
\title{Study of Holon Pair Condensation Based on the U(1) Slave Boson Approach to a Modified t-J Hamiltonian}
\author{Sung-Sik Lee and Sung-Ho Suck Salk}
\address{Department of Physics,\\
Pohang University of Science and Technology,\\
Pohang, Kyoungbuk, Korea 790-784}
\date{\today}

\maketitle

\begin{abstract}
In order to obtain a reasonable phase diagram involving hole-pair condensation for hole-doped cuprate systems, we propose a modified t-J Hamiltonian in which a Coulomb repulsion term between nearest neighbor holes is considered.
We show from the investigation of free energy that holon pair condensation can occur as a result of symmetry breaking.
In addition, we introduce a possibility of supersymmetry condition at $T=0K$ in the intermediate region between the antiferromagnetic phase and the superconducting phase. 
\end{abstract}
\pacs{PACS numbers: 74.25.-q, 74.25.Dw, 74.90.+n}

\begin{multicols}{2}

\newpage
The U(1) slave-boson approach to the t-J Hamiltonian is often employed for the study of high $T_{c}$ cuprates.
Earlier Kotliar and Liu studied a possibility of d-wave superconductivity as a result of d-wave spin singlet pairing and single boson condensation\cite{KOTLIAR}.
In their study, they introduced the hole pairing order parameter as the product of spinon and holon pair order parameters, i.e. $< c_{i \uparrow} c_{j \downarrow} > = < f_{i \uparrow} f_{j \downarrow} > < b_i^{\dagger} b_j^{\dagger}>$ in the mean field approximation\cite{KOTLIAR}.
Most recently, Wen and Lee proposed an SU(2) slave-boson theory of the t-J Hamiltonian and could not determine whether single boson condensation or boson pair condensation is favored for the superconducting phase\cite{WEN}.
From the exact diagonalization study of the t-J Hamiltonian, Riera and Dagotto stressed that intersite hole-hole repulsion should be introduced into the t-J Hamiltonian in order to properly account for the binding energy of the hole pair\cite{DAGOTTO98}.
In the present study, by paying attention to the hole-doped cuprate systems we present a possibility of holon pair condensation and present a phase diagram which exhibits the spin gap phase and the superconducting phase based on a modified t-J Hamiltonian.
In addition, we discuss a possibility of supersymmetry condition in an intermediate region between the antiferromagnetic phase and the superconducting phase at $T=0K$.

Satisfying the local U(1) gauge slave-boson representation\cite{KOTLIAR},\cite{UBBENS}, we write the following modified t-J Hamiltonian with the inclusion of an intersite hole-hole repulsion term(the third term) below,
\begin{eqnarray}
H & = & -t\sum_{<i,j>}(f_{i\sigma}^{\dagger}f_{j\sigma}b_{j}^{\dagger}b_{i} + c.c.) \nonumber \\
&& +J\sum_{<i,j>}({\bf S}_{i} \cdot {\bf S}_{j} - \frac{1}{4}n_{i}n_{j}) + 
V \sum_{<i,j>} b_{i}^{\dagger}b_{j}^{\dagger}b_{i}b_{j} \nonumber \\
&& - \mu_{0}\sum_{i,\sigma} f_{i\sigma}^{\dagger}f_{i\sigma} + i\sum_{i} \lambda_{i}(f_{i\sigma}^{\dagger}f_{i\sigma}+b_{i}^{\dagger}b_{i} -1).
\label{eq:tjmodel}
\end{eqnarray}
Here $f_{i\sigma}(f_{i\sigma}^{\dagger})$ is the spinon annihilation(creation) operator; $b_{i}(b_{i}^{\dagger})$, the holon annihilation(creation) operator; and $n_{i}$, the electron number operator at site $i$. 
$V$ in the third term is the strength of intersite hole-hole repulsion.
$\lambda_{i}$ is the Lagrangian multiplier to enforce local single occupancy constraint.
${\bf S}_{i}$ is the electron spin operator at site $i$, ${\bf S}_{i}=\frac{1}{2}f_{i\alpha}^{\dagger} \bbox{\sigma}_{\alpha \beta}f_{i\beta}$ with $\bbox{\sigma}_{\alpha \beta}$, the Pauli spin matrix element.
Unlike some other studies, in the present study we explicitly consider the intersite interaction term, $n_{i}n_{j}$ in Eq.(\ref{eq:tjmodel}) above.

The Heisenberg term in Eq.(\ref{eq:tjmodel}) can be decoupled into Hatree-Fock-Bogoliubov channels in association with the direct, exchange and pairing interactions of spinons.
By introducing corresponding Hubbard-Stratonovich fields, $\bbox{\rho}_{i}^{f}$, $\chi_{ji}$ and $\Delta_{ji}^{f}$\cite{UBBENS} and using $n_{i}=1-b_{i}^{\dagger}b_{i}$(with the neglect of double occupancy), we obtain the effective Hamiltonian from Eq.(\ref{eq:tjmodel}),
\begin{eqnarray}
H_{eff}  & = &  \frac{3J}{8} \sum_{<i,j>} \Bigl[ |\Delta_{ji}^{f}|^{2} + |\chi_{ji}|^{2} - \chi_{ji}^{*} \Bigl( f_{j\sigma}^{\dagger}f_{i\sigma}+\frac{8t}{3J}b_{j}^{\dagger}b_{i} \Bigr) -c.c. \nonumber \\
&& + n_{i} -\Delta_{ji}^{f*}(f_{j\uparrow}f_{i\downarrow}-f_{j\downarrow}f_{i\uparrow})-c.c. \Bigr] + \nonumber \\ 
&& + \frac{J}{2}\sum_{<i,j>}\Bigl( |\bbox{\rho}_{i}^{f}|^{2} - \sum_{k=1}^{3} (\rho_{i}^{f})^{k}(f_{j}^{\dagger}\sigma^{k}f_{j}) \Bigr) + \nonumber \\
&& -\mu_{0}\sum_{i}f_{i\sigma}^{\dagger}f_{i\sigma} -i\sum_{i}\lambda_{i}(f_{i\sigma}^{\dagger}f_{i\sigma}+b_{i}^{\dagger}b_{i}-1) + \nonumber \\
&& + \frac{8t^{2}}{3J}\sum_{<i,j>}(b_{j}^{\dagger}b_{i})(b_{i}^{\dagger}b_{j}) + V \sum_{<i,j>} b_{i}^{\dagger}b_{j}^{\dagger}b_{i}b_{j} + \nonumber \\
&& -\frac{J}{4}\sum_{<i,j>}b_{i}^{\dagger}b_{j}^{\dagger}b_{i}b_{j} + J\sum_{i}b_{i}^{\dagger}b_{i} - \frac{NJ}{2}. 
\label{eq:mf_hamiltonian1}
\end{eqnarray}

It is noted that the term, $\frac{8t^{2}}{3J} \sum_{<i,j>}(b_{j}^{\dagger}b_{i})(b_{i}^{\dagger}b_{j})$ in the expression above represents the exchange interaction(Fock) channel.
The last three terms come from the term, $-\frac{J}{4}\sum_{<i,j>} n_i n_j$ ($=  - \frac{J}{4}\sum_{<i,j>}b_{i}^{\dagger}b_{j}^{\dagger}b_{i}b_{j} + \frac{J}{2}\sum_{i}b_{i}^{\dagger}b_{i} -\frac{NJ}{2}$ in the slave-boson representation) in Eq.(\ref{eq:tjmodel}).
The first term here shows the contribution of attractive interaction between holons in the system of antiferromagnetically correlated systems.
Depending on emphasis, various forms of holon interaction potentials are mapped from the t-J Hamiltonian\cite{KOTLIAR}-\cite{SCHMELTZER}.
In our case we consider the decomposition of the holon interaction potential into the direct, exchange and pairing channels\cite{DECOMPOSITION}.
The contribution of the direct interaction term will be embedded in the effective holon chemical potential, $\mu^b_{i}$.
The exchange channel involves a large repulsive interaction of order $U \approx \frac{8t^2}{3J}$, as is seen from $\frac{8t^{2}}{3J}\sum_{<i,j>}(b_{j}^{\dagger}b_{i})(b_{i}^{\dagger}b_{j})$ in Eq.(\ref{eq:mf_hamiltonian1}) and is not readily accessible.
We obtain for the hole pairing channel,
\begin{eqnarray}
\displaystyle
&& e^{\frac{J_v}{4}\sum_{<i,j>} b_{i}^{\dagger}b_{j}^{\dagger}b_{i}b_{j}} \nonumber \\
&& \propto  \int \prod_{<i,j>} d\Delta_{ji}^{b*} d\Delta_{ji}^{b} e^{ -\frac{J_v}{4} \sum_{<i,j>} \Bigl[ |\Delta_{ji}^{b}|^{2} - \Delta_{ji}^{b*} (b_{j}b_{i}) - c.c \Bigr] }, 
\label{eq:holon_Hubbard_Stratonovich}
\end{eqnarray}
with $J_v=J-4V$.
Here $\Delta_b$ is the Hubbard-Stratonovich field for the holon pairing channel.
$J_v$ represents an effective intersite hole-hole attractive coupling.

By allowing the saddle point approximation, we obtain the mean field Hamiltonian from Eqs.(\ref{eq:mf_hamiltonian1}) and (\ref{eq:holon_Hubbard_Stratonovich}), 
\begin{eqnarray}
\lefteqn{H^{MF}=\sum_{<i,j>}\Bigl[ \frac{3J}{8} \Bigl( |\Delta_{ji}^{f}|^{2} + |\chi_{ji}|^{2} \Bigr) + \frac{J_v}{4} |\Delta_{ji}^{b}|^{2} \Bigr] +  }  \nonumber \\
& & -\frac{3J}{8} \sum_{<i,j>} \Bigl[ \Delta_{ji}^{f*} (f_{j\uparrow}f_{i\downarrow}-f_{j\downarrow}f_{i\uparrow}) + c.c. \Bigr] + \nonumber \\
&& -\frac{3J}{8} \sum_{<i,j>} \Bigl[ \chi_{ji}^{*} (f_{j\sigma}^{\dagger}f_{i\sigma}) + c.c. \Bigr]  + \nonumber \\
&& + \frac{J}{2}\sum_{<i,j>}\Bigl( |\bbox{\rho}_{i}^{f}|^{2} - \sum_{k=1}^{3} (\rho_{i}^{f})^{k}(f_{j}^{\dagger}\sigma^{k}f_{j}) \Bigr) - \sum_{i,\sigma} \mu^{f}_{i} (f_{i\sigma}^{\dagger} f_{i\sigma})  + \nonumber \\ 
& & -t \sum_{<i,j>} \Bigl[ \chi_{ji}^{*}(b_{j}^{\dagger}b_{i}) + c.c.  \Bigr]
-\frac{J_v}{4} \sum_{<i,j>} \Bigl[ \Delta_{ji}^{b*} (b_{i}b_{j}) + c.c. \Bigr]  + \nonumber \\
&& -\sum_{i} \mu_{i}^{b} ( b_{i}^{\dagger}b_{i} ),
\label{eq:mf_hamiltonian2}
\end{eqnarray}
where $\chi_{ji}= < f_{j\sigma}^{\dagger}f_{i\sigma} + \frac{8t}{3J} b_{j}^{\dagger}b_{i}>$, $\Delta_{ji}^{f}=< f_{j\uparrow}f_{i\downarrow}-f_{j\downarrow}f_{i\uparrow} >$, $\Delta_{ji}^{b} = <b_{j}b_{i}>$, $\bbox{\rho}_{i}^{f}=<{\bf S}_{i}>$,  $\mu_{i}^{f}=\mu_{0} + i\lambda_{i}-3J/4$ and $\mu_{i}^{b}=i\lambda_{i}-\frac{5J}{4}+V$.
Later $\bbox{\rho}_{i}^{f}$ will be taken to be $0$ as in Ref.\cite{UBBENS}.
For simplicity we allow uniform(site-independent) chemical potentials, $\mu^{f}_{i}=\mu^{f}$ and $\mu^{b}_{i}=\mu^{b}$.

Following Ubbens and Lee\cite{UBBENS} we now introduce the flux phase, $\chi_{ji}=\chi e^{\pm i\theta }$
where the sign $+(-)$ is along(against) the arrow indicated in Fig.1, and the pairing order parameters, $ \Delta_{ji}^{f}=\Delta_fe^{\pm i\tau^{f}} \mbox{ and } \Delta_{ji}^{b}=\Delta_be^{\pm i\tau^{b}}$,
where the sign $+(-)$ is for the ${\bf ij}$ link parallel to $\hat x$ ($\hat y$).
Here $\Delta_b$, $\Delta_f$ and $\chi$ denote the absolute magnitudes.
From now on the subscripts $i$ and $j$ will be deleted to allow uniform hopping and pairing order parameters.
We introduce the Bogoliubov-Valatin transformation following the momentum space representation of the mean field Hamiltonian $H^{MF}$ in Eq.(\ref{eq:mf_hamiltonian2}).
The resulting diagonalized Hamiltonian is,
\begin{eqnarray}
H^{MF} & = & \frac{3NJ}{4} \Bigl( (\Delta_f)^{2} + \chi^{2} \Bigr) + \sum_{k, s=\pm 1}^{'} E_{ks}^{f}(\alpha_{ks}^{\dagger}\alpha_{ks} - \beta_{ks}\beta_{ks}^{\dagger})  \nonumber \\
&& + \frac{NJ_v}{2} (\Delta_b)^{2} + \sum_{k, s=\pm 1}^{'} \{ E_{ks}^{b} (h_{ks}^{\dagger} h_{ks} + \frac{1}{2}) - \frac{\epsilon_{ks}^{b}}{2} \}  \nonumber \\
&& - N\mu^{f} + \frac{N\mu^{b}}{2}.
\label{eq:diagonalized_hamiltonian}
\end{eqnarray}
Here $E_{ks}^{f}$ and $E_{ks}^{b}$ are the quasiparticle energies:
\begin{equation}
E_{ks}^{f}  =  \sqrt{(\epsilon_{ks}^{f}-\mu^{f})^{2} + \Bigl( \frac{3J}{4} \xi_{k}(\tau^{f}) \Delta_f \Bigr)^{2}}, 
\label{eq:spinon_energy}
\end{equation}
for spinons and
\begin{equation}
E_{ks}^{b}  =  \sqrt{(\epsilon_{ks}^{b}-\mu^{b})^{2} - \Bigl( \frac{J_v}{2}\xi_{k}(\tau^{b}) \Delta_b \Bigr)^{2}}, 
\label{eq:holon_energy}
\end{equation}
for holons.
Here the symbol definitions are, for $\phi=\theta$, $\tau^{f}$ or $\tau^{b}$
\begin{eqnarray}
\xi_{k}(\phi) & = & \sqrt{ \gamma_{k}^{2} \cos^{2} \phi + \varphi_{k}^{2} \sin^{2} \phi }, \label{eq:xi},  \\
\epsilon_{ks}^{f} & = & \frac{3}{4}Js\chi \xi_{k}(\theta), \\
\epsilon_{ks}^{b} & = & 2ts \chi \xi_{k}(\theta), 
\end{eqnarray}
where $ \gamma_{k} = (\cos k_{x} + \cos k_{y})$, $\varphi_{k} = ( \cos k_{x} - \cos k_{y})$.
Here $\epsilon_{ks}^{f}$ and $\epsilon_{ks}^{b}$ are the quasiparticle energies for spinons and holons respectively in the absence of pairing, $\Delta_f=\Delta_b=0$.
$\alpha_{ks}( \alpha_{ks}^{\dagger})$ and $\beta_{ks}(\beta_{ks}^{\dagger})$ are the annihilation(creation) operators for the spinon quasiparticles, 'quasi-spinons' of spin up and spin down respectively, and $h_{ks}(h_{ks}^{\dagger})$, the annihilation(creation) operators of holon quasiparticle, 'quasi-holons'.
$\sum_{k}^{'}$ denotes a sum over half of the Brillouin zone.

From the diagonalized Hamiltonian in Eq.(\ref{eq:diagonalized_hamiltonian}), we readily obtain the total free energy, 
\begin{eqnarray}
&& F  =  \frac{3NJ}{4} \Bigl( (\Delta_f)^{2} + \chi^{2} \Bigr) 
- 2k_{B}T \sum_{k,s=\pm 1}^{'} ln [ \cosh (\beta E_{ks}^{f}/2) ] + \nonumber \\
&& - N\mu^{f} - 2Nk_{B}Tln2  + \frac{NJ_v}{2} (\Delta_b)^{2}  + \nonumber \\
&& + k_{B}T \sum_{k,s=\pm 1}^{'} ln [1 - e^{-\beta E_{ks}^{b}}] + \sum_{k,s=\pm 1}^{'} \frac{ E_{ks}^{b}-\epsilon_{ks}^{b}}{2} + \frac{N\mu^{b}}{2}.
\label{eq:free_energy}
\end{eqnarray}
By minimizing the above free energy, the amplitudes of the order parameters($\chi$, $\Delta_f$ and $\Delta_b$) can be obtained as a function of temperature and doping rate.
The chemical potentials at finite temperature can be determined from $\frac{\partial F}{\partial \mu^{b}} = -N\delta$ for the holons(quasi-holons) and from $\frac{\partial F}{\partial \mu^{f}} = -N(1-\delta)$ for the spinons(quasi-spinons).
From Eq.(\ref{eq:free_energy}) (in association with Eqs. (\ref{eq:spinon_energy}) through (\ref{eq:xi})), we note that the free energy is identical between the two sets of the phases of the hopping order parameter($\chi$) and the pairing order parameters($\Delta_f$ and $\Delta_b$): (1) $\theta=0$, $\tau^{f}=\pi/2$, $\tau^{b}=0$ and (2) $\theta=\pi/2$, $\tau^{f}=0$, $\tau^{b}=\pi/2$ (see Fig.1 for their notations). 
The first set refers to the d-wave spinon pairing and the s-wave holon pairing and the second set, the s-wave spinon pairing and the d-wave holon pairing.
In the present study we consider the first set in accordance with the experimentally observed d-wave spin gap phase\cite{YASUOKA}\cite{SHEN}.

Using Eq.(\ref{eq:free_energy}), we obtain  
\end{multicols}{2}
\vspace{-0.5cm}
\begin{eqnarray}
-\frac{\partial F}{\partial \mu^{b}} &=& \sum_{k,s=\pm 1}^{'} \Bigl[ \frac{1}{e^{\beta E_{ks}^{b}}-1} \frac{\epsilon_{ks}^{b}-\mu^{b}}{E_{ks}^{b}} + \frac{\epsilon_{ks}^{b}-\mu^{b}-E_{ks}^{b}}{2E_{ks}^{b}} \Bigr] = N \delta \label{eq:d_mu_F}, \\
\frac{\partial F}{\partial \Delta_b} &=& J_v \Delta_b \Bigl[ N - \sum_{k,s=\pm 1}^{'} \Bigl( \frac{1}{e^{\beta E_{ks}^{b}}-1}+ \frac{1}{2} \Bigr) \frac{J_v \xi_{k}(\tau^{b})^{2}}{4E_{ks}^{b}} \Bigr]=0 \label{eq:d_delta_F}.
\end{eqnarray}
\vspace{-0.5cm}
\begin{multicols}{2}
It is gratifying to find that for the case of no holon pairing, that is, $\Delta_b=0$, Eq.(\ref{eq:d_mu_F}) leads to the satisfactory Bose-Einstein statistical relation $\sum_{k,s=\pm 1}^{'} \frac{1}{e^{\beta (\epsilon_{ks}^{b}-\mu^{b})}-1} = N \delta$ for free holons(quasi-holons).
The free energy of the boson(holon pair) sector is from Eq.(\ref{eq:free_energy}) above, 
\begin{eqnarray}
F^{b} & = & \frac{NJ_v}{2} (\Delta_b)^{2} + k_{B}T \sum_{k,s=\pm 1}^{'} ln [1 - e^{-\beta E_{ks}^{b}}] \nonumber \\
& + & \sum_{k,s=\pm 1}^{'} \frac{ E_{ks}^{b}-\epsilon_{ks}^{b}}{2} + \frac{N\mu^{b}}{2}.
\label{eq:holon_free_energy}
\end{eqnarray}
This free energy is an even function of the complex holon pair order parameter, $\Delta_{ji}^{b}=\Delta_b e^{\pm i\tau^{b}}$, that is, $F^{b}(\Delta_b,\tau^{b}) = F^{b}(\Delta_b,\tau^{b}+\pi)$, as can be verified from Eq.(\ref{eq:holon_free_energy}) in association with Eq.(\ref{eq:holon_energy}).
We readily see from Eq.(\ref{eq:d_delta_F}) that $\frac{\partial F}{\partial \Delta_b}=0$ for $\Delta_b=0$.
This indicates that the free energy as an even function has a symmetry at $\Delta_b=0$.

It is now of great interest to see if this symmetry can be spontaneously broken, by investigating a possible existence of a 'Mexican hat' form of free energy, that is, the condition of $\left. \frac{\partial^{2} F}{\partial \Delta_b^{2}} \right|_{\Delta_b=0}<0$ at a critical temperature and a minimum of free energy at $\Delta_b \neq 0$.
We obtain from Eq.(\ref{eq:d_delta_F}),
\begin{eqnarray}
 &&\left. \frac{\partial^{2} F^b}{\partial \Delta_b^{2}} \right|_{\Delta_b=0} = \nonumber \\
 && J_v \Bigl[ N - \sum_{k,s=\pm 1}^{'} \Bigl( \frac{1}{e^{\beta (\epsilon_{ks}^{b}-\mu^{b})}-1}+ \frac{1}{2} \Bigr) \frac{J_v \xi_{k}(\tau^{b})^{2}}{4(\epsilon_{ks}^{b}-\mu^{b})}
\Bigr].
\label{eq:2nd_derivative}
\end{eqnarray}
Here we examine the second term in the bracket of Eqs.(\ref{eq:2nd_derivative}).
$\xi_k(\tau^b)$ is independent of temperature with a bounded value, $0 \leq \xi_{k}(\tau^{b}) \leq 2$.
As temperature increases, the denominator $\epsilon_{ks}^{b}-\mu^{b}$ increases due to the increased population of high lying levels $\epsilon_{ks}^{b}$.
As a result this allows a possibility of $\left. \frac{\partial^{2} F}{\partial \Delta_b^{2}} \right|_{\Delta_b=0} > 0$.
On the other hand, as temperature decreases, $\epsilon_{ks}^{b}-\mu^{b}$ decreases due to the increased population of low lying excitations $\epsilon_{ks}^{b}$, thus allowing a possibility of $\left. \frac{\partial^{2} F}{\partial \Delta_b^{2}}\right|_{\Delta_b=0} < 0$ at a critical temperature.
This causes the instability of free holons against holon pairing at the critical temperature.
Further the computed holon pair free energy in Eq.(\ref{eq:holon_free_energy}) yields its minimum at a finite value of the holon pairing parameter, $\Delta_b \neq 0$.
Thus we find that the 'Mexican hat' form of free energy exists to cause holon pair condensation as a result of symmetry breaking.

Currently it is of great interest to predict observed phase diagrams(Fig.2.,\cite{YASUOKA}) involving both the spin gap(pseudo gap) and superconducting phases.
In Fig.3 we display a computed phase diagram obtained from the minimization of the free energy in Eq.(\ref{eq:free_energy}) with respect to the order parameters, $\chi$, $\Delta_f$ and $\Delta_b$. 
Encouragingly, with the inclusion of the Coulomb repulsion term (denoted as $T^b_{V \neq 0}$) we were able to obtain at $J=0.2t$ a phase diagram with reasonable values of optimal doping rate and critical temperature, as is shown in Fig.3.
This is in qualitative agreement with observation\cite{YASUOKA}.
Here $V=4.99 \times 10^{-2}t$ was used for the hole-hole repulsion energy.
However, with the neglect of this Coulomb repulsion term, i.e. $V=0$ (denoted as $T^b_{V=0}$), it was impossible to fit the experimentally observed phase diagram.

Thus far we examined the free energy at finite temperatures, $T \neq 0$ by paying attention to the spin gap and superconducting phases.
It is now of interest to examine at $T=0K$ the intermediate region between the antiferromagnetic phase and the superconducting phase.
Choosing the case of equality between the spinon quasiparticle energy and the holon quasiparticle energy, that is, $E^f_{ks}=E^b_{ks}$, Eq.(\ref{eq:diagonalized_hamiltonian}) leads to
\begin{equation}
H_{SUSY} = \sum_{k, s=\pm 1}^{'} E_{ks}(g_{ks}^{\dagger}g_{ks} + h_{ks}^{\dagger} h_{ks} ),
\label{eq:susy_hamiltonian}
\end{equation}
with $g_{ks}=\alpha_{ks}$ or $\beta_{ks}$.
The equality, $E^f_{ks}=E^b_{ks}$ is satisfied under the condition of $ \Delta_f  =  \Delta_b = 0$, $\chi  =  0$ and  $\mu^{f}  =  \mu^{b}$.
The above Hamiltonian is now realized as a SUSY(supersymmetry) Hamiltonian.
This is because the SUSY algebra $ \{ Q,Q \}  =  H_{SUSY}$ is satisfied with the supercharge operator, $ Q  = \sum_{k, s=\pm 1}^{'} \sqrt{\frac{E_{ks}}{2}} \Bigl( g_{ks}^{\dagger} h_{ks} + h_{ks}^{\dagger} g_{ks} \Bigr)$\cite{LAHIRI}.
It is of great interest to experimentally verify the SUSY condition at $T=0K$, i.e., $ \Delta_f  =  \Delta_b = 0$, $\chi  =  0$ and  $\mu^{f}  =  \mu^{b}$.
The SUSY condition may be observed in the intermediate doping region (i.e. $\delta_{A.F.} < \delta < \delta_{A.F.}$ in Fig.2) between the antiferromagnetic phase and the superconducting phase at $T=0K$.

In the present study, by proposing a modified t-J Hamiltonian with the inclusion of intersite Coulomb repulsion term, we were able to obtain a reasonable phase diagram for the hole-doped cuprate systems. 
We found that the U(1) slave-boson theory of the t-J Hamiltonian in its original form (with the neglect of the repulsion term) can not reproduce the superconducting phase for the hole-doped cuprate systems. 
We showed from the investigation of the free energy in Eq.(\ref{eq:free_energy}) that holon pair condensation can occur as a result of symmetry breaking.
The d-wave hole pairing can be realized from the mean field account of $<c_{i \uparrow}c_{j \downarrow}> = <f_{i \uparrow}f_{j \downarrow}> <b_i^{\dagger} b_j^{\dagger}>$.
It can arise with the s-wave holon pairing ($<b_i^{\dagger} b_j^{\dagger}>$) under the condition of the d-wave singlet pairing ($<f_{i \uparrow}f_{j \downarrow}>$).
Finally, we introduced a possibility of supersymmetry condition at $T=0$ which may exist in the intermediate region between the antiferromagnetic phase and the superconducting phase. 
It is of great interest to see its experimental verification in the future.

One of the authors(SHSS) acknowledges the generous supports of Korea Ministry of Education(BSRI-97) and the Center for Molecular Science at Korea Advanced Institute of Science and Technology.
He is also grateful to Professor R. B. Laughlin for stimulating discussions on supersymmetry.
We thank Mr. Tae-Hyoung Gimm for helpful discussions.

\begin{minipage}[c]{8cm}
\begin{figure}
\hspace{1.5cm}
\epsfig{file=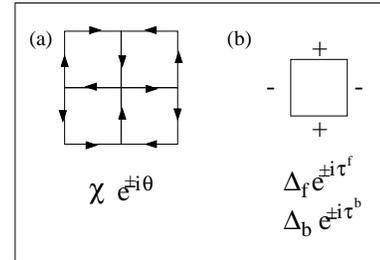, angle=270,width=5cm}
\label{fig:flux}
\caption{ Phases of hopping and pairing order parameters. (a) $+\theta(-\theta)$ along(against) the arrow on the chain. (b) $+\tau(-\tau)$ in the chain of $\hat x(\hat y)$ direction(Ref.[4]).}
\end{figure}
\end{minipage}

\begin{minipage}[c]{8cm}
\begin{figure}
\hspace{1.5cm}
\epsfig{file=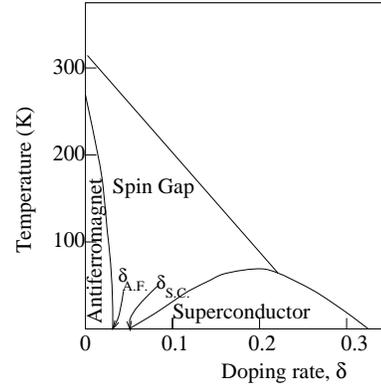,width=5cm}
\label{fig:sc_phase}
\caption{Schematic phase diagram of $La_{2-\delta}Sr_{\delta}CuO_{4}$(Ref.[7]).}
 \end{figure}
 \end{minipage}

 \begin{minipage}[c]{8cm}
 \begin{figure}
 \hspace{1.5cm}
 \epsfig{file=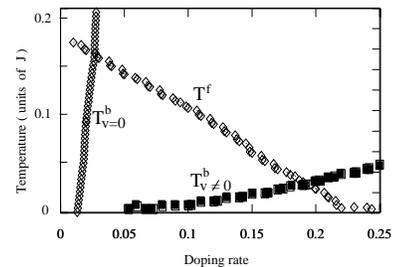,width=5cm}
 \label{fig:phase}
 \caption{ Computed phase diagram.  $T^b_{V=0}$ is the holon pair condensation temperature for $V=0$ and $T^b_{V \neq 0}$, the holon pair condensation temperature for $V=4.99 \times 10^{-2}t$.}
 \end{figure}
 \end{minipage}

\end{multicols}
\end{document}